\documentclass{osa-article}

\journal{boe}

\articletype{Research Article}

\begin{document}

\title{Extracting single- and multiple-scattering components in laser speckle contrast imaging of tissue blood flow}

\author{Yifan Zhang,\authormark{1} Cheng Wang,\authormark{2} Shanbao Tong,\authormark{1} and Peng Miao\authormark{1,*}}

\address{\authormark{1}School of Biomedical Engineering, Shanghai Jiao Tong University, Shanghai 200240 China\\
\authormark{2}School of Mathematical Sciences, Shanghai Jiao Tong University, Shanghai 200240 China}

\email{\authormark{*}pengmiao@sjtu.edu.cn} 



\begin{abstract}
Random matrix theory provides new insights into multiple scattering in random media. In a recent study, we demonstrated the statistical separation of single- and multiple-scattering components based on a Wishart random matrix. The first- and second-order moments were estimated through a Wishart random matrix constructed using dynamically-backscattered speckle images. In this study, this new strategy was applied to laser speckle contrast imaging (LSCI) of \textit{in vivo} blood flow. The random matrix-based method was adapted and parameterized using electric field Monte Carlo simulations and \textit{in vitro} blood flow phantom experiments. The new method was further applied in \textit{in vivo} experiments, demonstrating the benefits of separating the single- and multiple-scattering components, and was compared with the traditional temporal LASCA method. More specifically, the new method captures stimulus-induced functional changes in blood flow and tissue perfusion in the superficial and deeper layers. The new method extends the ability of LSCI to image functional and pathological changes.
\end{abstract}

\section{Introduction}
Blood flow and tissue perfusion are important physiological and pathological parameters in clinical diagnostics. Laser speckle contrast imaging provides full-field, high-resolution, real-time imaging of blood flow and tissue perfusion. As a label-free imaging modality, it measures relative blood flow based on the contrast analysis of backscattered speckle images\cite{dunn2012laser}. During the last two decades, laser speckle contrast imaging has been used to image the cerebral cortex, retina, skin, mesentery, bone joints, and other structures\cite{boas2010laser,senarathna2013laser,briers1982retinal}. LSCI is usually used in reflective imaging geometry with full-field illumination in continuous wave (CW) mode\cite{davis2014imaging,briers2013laser}. More specifically, it uses the coherent light, \emph{e.g.}, a near-infrared laser, to illuminate the tissue surface, gathers the backscattered speckle images and calculates the contrast image to represent the relative blood flow maps. The contrast value in each pixel of the contrast image is defined as the ratio of the standard deviation to the mean intensity\cite{boas1997spatially,vaz2016laser}, i.e. $K=$ $\sigma / \mu$. This ratio is theoretically related to the blood flow velocity $v$ through Eq.~(\ref{eq:one})\cite{briers2001laser}:
\begin{equation}
K^{2}=\beta\left\{\frac{\tau_{c}}{T}+\frac{\tau_{c}^{2}} {2 T^{2}} \left[e^{-2 T / \tau_{c}}-1\right]\right\}
\label{eq:one}.
\end{equation}
where $\beta$ is a constant parameter that accounts for the number of speckles in each pixel area,  $T$ is the exposure time of the camera, and $\tau_{c}$ is the decorrelation time, which is inversely proportional to the blood flow velocity $v$.

Since LSCI utilizes full-field CW illumination, the Brownian motion of scatterers in tissue causes temporal intensity fluctuations in the recorded speckle images \cite{volpe2014brownian,maret1987multiple}. The ordered blood flow in tissue blurs such dynamics and reduces contrast values. Compared to pointwise or line-scan illumination, full-field illumination causes the detected intensity at each pixel to contain more multiple-scattering trajectories. Singly- and multiply-backscattered light coexist in the speckle patterns. The singly-backscattered light enters the tissue and is backscattered once. This component primarily penetrates the superficial layer of tissue $\left(< \mathrm{l}_{\mathrm{sa}}/2\right)$. The superposition of both components, \emph{e.g.}, the recorded speckle images, always suppresses some of the details in both components.

Biological tissues are complex forms of random media. Coherent waves propagating through a random medium undergo multiple scattering processes with interference phenomena. For transmissive imaging, optically thick tissue is turbid or opaque because of strong multiple scattering. Even in semitransparent tissues, no clear structures can be revealed when the transmitted light is multiply scattered. Thus, for \emph{in vivo} applications, reflective imaging or detection is preferred. To describe light propagation, a traditional radiative transport equation (RTE)-based model can be used, but these models do not account for interference phenomena\cite{mishchenko2006multiple}. However, the speckle images captured by LSCI have interference origins, \emph{i.e.},  the coherent addition of backscattered light with random phasors due to different paths. Random matrix theory (RMT) has been successfully used to describe the behavior of coherent wave multiple scattering in random media\cite{beenakker1997random, fyodorov1997statistics, aubry2009random}. Light focusing and target detection have been achieved in turbid and opaque samples using acoustic and light waves\cite{popoff2010measuring, angelsen2000ultrasound, stergiopoulos2003advanced}. We recently established a random matrix (RM)-based description of dynamically-backscattered speckle patterns derived from trajectory perturbations due to the Brownian motion of scatters \cite{miao2021random}. A strategy for separating the single and multiple scattering components in backscattered light has also been developed.

In most cases, separating the single- and multiple-scattering components is difficult in reflective imaging. Pulse illumination with time gating\cite{kang2015imaging} (time of flight) and low-coherence light with coherence gating (OCT)\cite{huang1991optical} utilize the path length difference to filter out the multiple scattering component. A source-detector separation strategy can filter out the single scattering component (NIRS)\cite{jobsis1977noninvasive}. However, the traditional method cannot be used with full-field CW coherent illumination. Our RMT-based method facilitates the separation of single and multiple scattering components in \textit{in vivo} tissue imaging applications. In this study, we adapted and optimized the RMT-based method for LSCI to monitor tissue blood flow and functional changes.

\section{Theory}

Standard LSCI utilizes full-field illumination with a coherent NIR laser ($785 \mathrm{~nm}$ in this study) in $\mathrm{CW}$ mode. The backscattered speckle pattern is recorded by a monochronic camera with an exposure procedure. The gray levels in the recorded image $I\left[n_{1}, n_{2}\right], n_{1}=1 \cdots N_{1}, n_{2}=1 \cdots N_{2}$ are proportional to the light intensities. There are single and multiple scattering contributions in each pixel (entry), i.e., $I\left[n_{1}, n_{2}\right]=I_{S}\left[n_{1}, n_{2}\right]+I_{M}\left[n_{1}, n_{2}\right]$. Since multiple scattering contributes to the speckle pattern diffusively, the entries in $I_{M}$ follow the Gaussian distribution $\left(\sim \mathbb{N}\left(\mu_{M}, \sigma_{M}^{2}\right)\right)$. The intensity contribution from the single-scattering component follows a negative exponential distribution $(\sim \operatorname{Exp}(\lambda))$.

To obtain the dynamically-backscattered speckle images, sequential recordings were performed: $\left\{I_{i} \mid i=1 \cdots T\right\}$. Since LSCI uses a monochromic camera with a low frame rate ($50$ fps) and a short exposure time ($5 \mathrm{~ms}$), the interframe time interval is long enough to ensure the sampling is statistically independent. Brownian motion introduces dynamic fluctuations in recorded speckle images. For the multiple-scattering component, the temporally-independent sampling follows \emph{i.i.d.} realizations from $\mathbb{N}\left(\mu_{M}, \sigma_{M}^{2}\right)$. On the other hand, the single-scattering component produces relatively stable patterns due to the presence of in-phase paths that form low rank characteristics. Blood flow provides another source of dynamic effects in both the single and multiple scattering components. In the following theory development, we ignore the blood flow effect at first and revisit its effects in the contrast analysis.

For $T$ recorded speckle images $\left\{I_{i} \mid i=1 \cdots T\right\}$, each $I_{i}$ is reshaped to a column vector $h_{i}$, with entries $h_{i}[n], n=1 \cdots N, N=N_{1} \times N_{2}$ as the $\mathrm{i}^{\text {th }}$ column in the $N \times T$ temporal intensity random matrix $R$. Then, $R=R_{S}+R_{M}$, and we can further centralize $R$ by $\widehat{R}=R-\bar{R}$, where each entry in $\bar{R}$ is the mean value of the same row. We have $\hat{R}=\hat{R}_{S}+\widehat{R}_{M}$, where the subscripts $\quad S$ and $M$ denote the single and multiple scattering components, respectively.

We next investigated the spectral density of the Wishart random matrix, $E=\hat{R} \hat{R}^{\prime}$. The eigenvalues of $E$ are denoted by $\{s(i)\}$, with $i=1 \cdots N$ and $s(1) \geq s(2) \geq \cdots \geq s(N)$. The probability density of the eigenvalues can be calculated as $\rho(s) \triangleq \frac{1}{N} \sum_{i=1}^{N} \delta(s-s(i))$.

For the centralized multiple-scattering component, \emph{i.e.}, $\hat{R}_{M}$, the entries follow the \emph{i.i.d.} Gaussian distribution $\left(\sim \mathbb{N}\left(0, \sigma_{M}^{2}\right)\right)$. The eigenvalue density of $E_{M}$ is well described by the Mar$\rm \check{c}$enko-Pastur law with $T \geq N$ and $N,T \rightarrow \infty$\cite{marchenko1967distribution}:

\begin{equation}
\rho\left(s_{M}\right)=\frac{Q}{2 \pi \sigma_{M}^{2}} \frac{\sqrt{\left(s_{M+}-s_{M}\right)\left(s_{M}-s_{M-}\right)}}{s_{M}}
\label{eq:two}.
\end{equation}
where $Q=T/N$, $\{s_{M}(i)\}, i=1 \cdots N$ are the eigenvalues of $E_{M}$, and $s_{M \pm}=\sigma_{M}^{2}(1 \pm \sqrt{1 / Q})^{2}$.

When the eigenvalue density of $E_{M}$ has finite $4^{\text {th }}$ moments, we have the following convergence about the maximum and minimum eigenvalues $s_{M}(1)$\cite{bai1988note} and $s_{M}(N)$\cite{bai2008convergence}:

\begin{equation}
\left(\begin{array}{l}
s_{M}(1) \underset{T \rightarrow \infty}{\longrightarrow} s_{M+} \\
s_{M}(N) \underset{T \rightarrow \infty}{\longrightarrow} s_{M-}
\end{array} \right)
\label{eq:three}.
\end{equation}

Eq.~(\ref{eq:three}) provides the theoretical basis for estimating the intensity variance of the multiple scattering components using the extreme eigenvalues of $E_{M}$. In practice, only the eigenvalues of $E$ can be calculated, which are biased from those of $E_{M}$ due to single-scattering components. However, the low rank characteristics of single-scattering components limit their bias effects on the larger eigenvalues of $E$. The eigenvalues of $E_{M}$ and $E$ converge to each other as the eigenvalues decrease to their minimum values (here, the eigenvalues of $E_{s}$ are nearly zero). Loubaton and Vallet have proven the convergence of both minimal eigenvalues. Thus, we can use the minimal eigenvalue of $E$, i.e., $s(N)$, to estimate $\sigma_{M}^{2}$ :

\begin{equation}
s(N) \underset{T \rightarrow \infty}{\longrightarrow} \sigma_{M}^{2}(1 \pm \sqrt{1 / Q})^{2}
\label{eq:four}.
\end{equation}

Then, we can analyze the trace relation (Eq.~(\ref{eq:five})) and the sampling variance $\tilde{\sigma}^{2}=\operatorname{tr}\left(\tilde{R} \tilde{R}^{\prime}\right) / N T, \tilde{\sigma}_{S}^{2}=$ $\operatorname{tr}\left(\tilde{R}_{S} \tilde{R}_{S}^{\prime}\right) / N T, \tilde{\sigma}_{M}^{2}=\operatorname{tr}\left(\tilde{R}_{M} \tilde{R}_{M}^{\prime}\right) / N T$.

\begin{equation}
\operatorname{tr}\left(\tilde{R} \tilde{R}^{\prime}\right)=\operatorname{tr}\left(\tilde{R}_{S} \tilde{R}_{S}^{\prime}\right)+2 \operatorname{tr}\left(\tilde{R}_{S} \tilde{R}_{M}^{\prime}\right)+\operatorname{tr}\left(\tilde{R}_{M} \tilde{R}_{M}^{\prime}\right)
\label{eq:five}.
\end{equation}

Since the multiple-scattering component $\tilde{R}_{M}$ follows the MP law, we have $\tilde{\sigma}_{M}^{2} \rightarrow \sigma_{M}^{2}$ and $\tilde{R}_{M}[n,t] \sim N\left(0, \sigma_{M}^{2}\right) .$ Then, we have the commutative part $\operatorname{tr}\left(\tilde{R}_{S} \tilde{R}_{M}^{\prime}\right) / M T \sim \mathbb{N} \left(0, \tilde{\sigma}^{2} / N T\right) \approx 0$. Finally, we have Eq.~(\ref{eq:six}) to estimate $\sigma_{S}\left(\sigma_{S}^{2} \approx \tilde{\sigma}_{S}^{2}\right)$.

\begin{equation}
\tilde{\sigma}_{S}^{2} \approx \tilde{\sigma}^{2}-\tilde{\sigma}_{M}^{2}
\label{eq:six}.
\end{equation}

After obtaining $\sigma_{S}$, the corresponding mean intensity $\mu_{S}$ can be obtained since the exponential distribution has $\mu_{S}=\sigma_{S}$. Finally, the mean intensity can be found: $\mu_{M}=\mu-\mu_{S}$.

The effect of ordered flow can be modeled by an additional drift component in the original Brownian motion. This will blur the temporally-averaged speckle patterns. The blurring effect leads to a decrease in both $\sigma_{S}$ and $\sigma_{M}$. Traditionally, we calculate the contrast values $K=\sigma / \mu$ to quantify the blurring effect of ordered flow because the mean intensities are inhomogeneous in the full-field imaging area. This calculation is also used in the new method for calculating multiple-scattering contrast images: $K_{M}=\sigma_{M} / \mu_{M}$. For single-scattering components, the light path is concentrated in the superficial layer and changes in $\mu_{S}$ follow changes in $\sigma_{S}$ synchronously. Therefore, we can directly use the contrast $K_{S}=\sigma_{S}$ to indicate the relative flow velocity in the single-scattering contrast image. The relations among $K_{S}$, $K_{M}$ and the flow velocity $v$ have been validated using both electric field Monte Carlo simulations and \textit{in vitro} blood flow phantom experiments, as described below.

\section{Methods and experiments}
\subsection{Electric Field Monte Carlo (EMC) simulation}

Radiation transport equations (RTEs) provide a comprehensive description of light propagation in random media. In practice, analytic solutions of RTEs are difficult to obtain. As a numerical tool, Monte Carlo simulation has been developed to investigate light propagation; the results are consistent with RTE results\cite{kasen2006time}. BBoth RTEs and traditional Monte Carlo methods ignore interference effects in coherent light transportation. To simulate coherent propagation, changes in light fields need to be tracked at each scattering event along the light trajectories. Mie scattering theory analytically supports such tracking tasks. Xu and colleagues developed electric field Monte Carlo (EMC) simulations as a tool for investigating coherent propagation\cite{xu2004electric}. EMC simulations track field changes using Mie scattering formulas. The interference effects can be obtained by the coherent addition of the electric fields. In this study, we extended the EMC simulation program by realizing full-field coherent illumination and detection of speckle patterns.

To simulate blood flow in biological tissue, we developed a $400\mu m \times 400\mu m \times 400\mu m$ phantom that contained an aqueous solution of homogeneously and randomly distributed Mie scatterers (diameter $a=0.8 \mu m$, volume fraction 50\%). A vessel cylinder (calibre $r = 70 \mu m$) embedded close to the surface layer contained the same scatterers but with additional ordered flow motion. A matching boundary condition was used to simplify the numerical simulation. A coherent $800 \mathrm{~nm}$ linearly polarized plane wave illuminated the surface, yielding a size parameter $x=2.09$, a transport mean free path $l_{t}=2.066$, a refractive index $n=1.59$, and an anisotropic coefficient $g=0.685$. In each simulation, we launch $10^{8}$ as full-field incident light and recorded the position and electric field changes of each scattering event along all backscattered light trajectories. To obtain the steady-state output of CW mode illumination, all the backscattered path fields were added coherently, resulting in the single speckle image $I[n_1,n_2] = |\sum E_x[n_1,n_2]+ \sum E_y[n_1,n_2]|^2$. 

For each scattering event, the local coordinate system  $\left(\boldsymbol{e}_{1}, \boldsymbol{e}_{2}, \boldsymbol{q}\right)$ was rotated to $\left(\boldsymbol{e}_{1}^{\prime}, \boldsymbol{e}_{2}^{\prime}, \boldsymbol{q}^{\prime}\right)$ by applying the scattering angles $(\theta, \phi)$ (Eq.~(\ref{eq:seven})) sampled from the distribution of the normalized phase function $p(\theta, \phi)=F(\theta, \phi) / \pi Q_{s c a} x^{2}$, where $Q_{s c a}$ is the scattering efficiency, $x=k a$ is the size parameter, $a$ is the radius of the particle and $F(\theta, \phi)$ is the scattered light intensity propagating along direction $(\theta, \phi)$.

\begin{equation}
\left(\begin{array}{l}
\boldsymbol{e_{1}}^{\prime} \\
\boldsymbol{e_{2}}^{\prime} \\
\boldsymbol{q}^{\prime}
\end{array}\right)=A\left(\begin{array}{c}
\boldsymbol{e}_{1} \\
\boldsymbol{e}_{2} \\
\boldsymbol{q}
\end{array}\right)
\label{eq:seven}.
\end{equation}

with
$$
A=\left(\begin{array}{ccc}
\cos \theta \cos \phi & \cos \theta \sin \phi & -\sin \phi \\
-\sin \phi & \cos \phi & 0 \\
\sin \theta \cos \phi & \sin \theta \sin \phi & \cos \theta
\end{array}\right)
$$

Each scattering event scatters the incident electric field $\boldsymbol{E}=E_{1} \boldsymbol{e}_{\mathbf{1}}+E_{2} \boldsymbol{e}_{\mathbf{2}}$ to $\boldsymbol{E}^{\prime}=E_{1}^{\prime} \boldsymbol{e}_{1}^{\prime}+E_{2}^{\prime} \boldsymbol{e}_{2}^{\prime}$ through Eq.~(\ref{eq:eight}).

\begin{equation}
\left(\begin{array}{l}
E_{1}^{\prime} \\
E_{2}^{\prime}
\end{array}\right)=B\left(\begin{array}{l}
E_{1} \\
E_{2}
\end{array}\right)
\label{eq:eight}.
\end{equation}

with
$$
B=[F(\theta, \phi)]^{-1 / 2}\left(\begin{array}{cc}
S_{2} \cos \phi & S_{2} \sin \phi \\
-S_{1} \sin \phi & S_{1} \cos \phi
\end{array}\right)
$$
where $S_{1}$ and $S_{2}$ are the perpendicular and parallel electric fields projected to the scattering plane (spanned by $q$ and $q^{\prime}$, respectively).

To obtain the Brownian motion-produced dynamic speckle images, the positions in all trajectories were changed based on the random walk model, \emph{i.e.}, $\Delta r(\tau) \sim \mathbb{N} (0,\sqrt{6D_B\tau})$, where $D_B$ is the scatterer diffusion coefficient. Given the momentum relaxation time and configurational relaxation time, the time scale used in the simulation was set to $10^{-4} \sim 10^{-3} s$. The backscattered fields were recalculated to form a new speckle image. To simulate ordered motion (blood flow), additional constant drifts $v\tau$ were added to each $\Delta r(\tau)$. To estimate the linearity of the contrast values obtained from the single- and multiple-scattering components, $v \in \left[1,7] \mathrm{mm} / \mathrm{s}\right)$ was applied to the Brownian motion-induced trajectory changes inside the vessel cylinder. For each configuration, we averaged $10$ speckle images as one recorded speckle image obtained by camera exposure ($5 ms$ in this study). 

\subsection{\textit{In vitro} flow phantom experiment}

We then used an \textit{in vitro} flow phantom experiment to test the accuracy and linearity of the flow velocity estimation using the contrast values obtained from the separated single- and multiple-scattering components. Intralipid (IL) (Kabivitrum Inc., USA) solution was used as the Mie scattering random medium. The mean diameter of the lipid droplets was $0.7 \mu \mathrm{m}$, and its scattering coefficient $\mu_{\mathrm{s}}$ at $785 \mathrm{~nm}$ was approximately $2 \mathrm{~mm}^{-1}$ at 2\% concentration. The mean free path (MFP) was $0.5 \mathrm{~mm}$ at a 2\% concentration. We used a motorized pump syringe to control the flow of the IL solution inside a polyethylene tube (PE-50, outer diameter: $0.97 \mathrm{~mm}$; inner diameter: $0.58 \mathrm{~mm}$ ) at different velocities $(2 \mathrm{~mm} / \mathrm{s} \sim 10 \mathrm{~mm} / \mathrm{s})$.

A diode laser ($785 \mathrm{~nm}$, LP785-SF20, Thorlabs, USA) was used as the monochromic coherent light source. The laser beam was expanded by a diffuser and illuminated the surface of the phantoms (Fig.~\ref{fig:3}(a)). The reflected light (speckle pattern) was imaged by a monochrome 12-bit CCD camera (SCA640-70fm, Basler Scout, Germany) with a macro lens (AF-S DX Micro $40 \mathrm{~mm}$ f2.8, Nikon) focused on the surface of the phantom. The recorded speckle images were further processed to construct the hybrid Wishart RM and estimate the first and second moments of the single- and multiple-scattering components, respectively. Finally, the contrast values of the single- and multiple-scattering components were calculated using different window sizes to cover the tube area and compared with the true velocity for parameter optimization.

\subsection{\textit{In vivo} CBF imaging applications}
\subsubsection{Steady-state CBF imaging of rats}

All animal experimental procedures were performed using protocols approved by the Animal Care and Use Committee of Shanghai Jiao Tong University. Adult Wistar rats $(\sim 250 \mathrm{~g}$, female) were anesthetized with sodium pentobarbital (3ml/kg, IP) and placed in a stereotactic frame (David Kopf Instruments, Tujunga, CA, USA). A homeothermic blanket system was used to maintain the rectal temperature of the rats at $37^{\circ} \mathrm{C}$. After a midline incision on the scalp, a high-speed dental drill (Fine Science Tools Inc. North Vancouver, Canada) was used to thin a $5 \mathrm{~mm} \times 5 \mathrm{~mm}$ area centered $3.5 \mathrm{~mm}$ lateral to and $3 \mathrm{~mm}$ posterior to the bregma.

A 12-bit cooled monochrome CCD camera (Sensicam SVGA, Cooke, Michigan, USA) with a $60 \mathrm{~mm}$ f/2.8 macro lens (Nikon Inc., Melville, NY, USA) was used to record the laser speckle images $(1280 \times 1024$ pixels, $10 \mathrm{fps}$ ) under He-Ne laser illumination $(632.8 \mathrm{~nm}, 0.5 \mathrm{~mW}$, JDSU, Milpitas, California). A total of 30 speckle images were recorded. The single- and multiple-scattering components were separated after constructing the Wishart RM. After that, the contrast images of the single- and multiple-scattering components were calculated. The traditional tLASCA algorithm was also applied to the recorded speckle images to obtain a hybrid contrast image. The results using different temporal and spatial window sizes were calculated and compared for parameter optimization.

After LSCI, fluorescence imaging was performed on the rat to obtain ground truth images of the cortical vasculature. Under anesthesia, $200 \mathrm{~mL}$ of rhodamine-dextran tracer (Invitrogen, Carlsbad, California, USA) was injected into the bloodstream via the femoral vein. Fifteen minutes after the injection, the distribution of the rhodamine-dextran in the cerebral cortex was captured using the same camera but with a bandpass emission filter ($560 \pm 5 nm$) and under the fluorescence excitation of a $532 \mathrm{~nm}$ diode pumped solid-state (DPSS) laser ($5 \mathrm{~mW}$, Nd: YVO4KTP, Beam of Light Technologies, Clackamas, Oregon, USA). To enhance the SNR of the fluorescence signal, a total of 20 fluorescence emission images were averaged to obtain the final fluorescence image.

\subsubsection{Functional CBF imaging after electrical hind paw stimulation}

The surgical preparation was similar to the steady-state rat CBF imaging. After the thinned skull preparation, two needle electrodes were inserted subcutaneously into the rat’s right hind paw. A stimulator was used to inject the current stimulation trials. Each trial consisted of a baseline period ($5 s$), electrical stimulation ($10 s$, rectangular constant current pulses ($2.5 mA, 0.3 ms, 5 Hz$) and a recovery period ($300 s$). During the direct electrical stimulation experiment, the cranial window was illuminated by the He-Ne laser and imaged by the CCD camera to record the functional changes. The imaging ROI encompassed the cerebral cortex surrounding the electrode. Images were continuously recorded at $10 fps$ with an exposure time of $5 \mathrm{~ms}$. The optimal parameters were applied to process the functional CBF images. The contrast image series corresponding to both the single- and multiple-scattering components were obtained and analyzed for the CBF response to electrical stimulation. The imaging ROIs covered the contralateral somatosensory area of the rat cerebral cortex.

\begin{figure}[htbp]
\centering\includegraphics[width=0.95\textwidth]{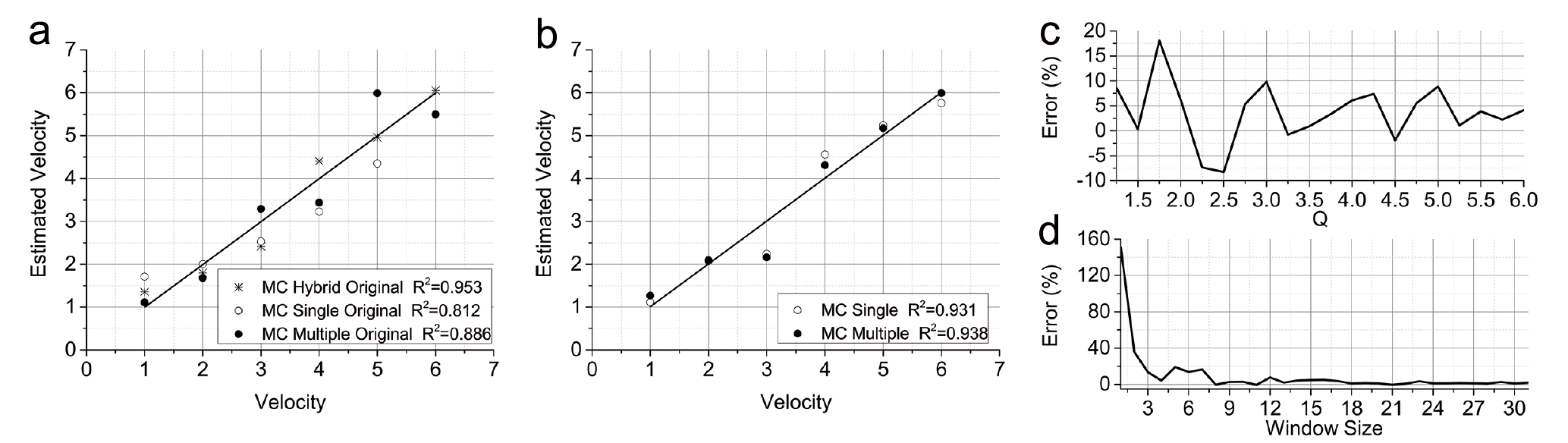}
\caption{\label{fig:1} EMC simulation results: (a) The calculated velocities from the single-scattering trajectory, multiple-scattering trajectory and hybrid trajectory had linear relationships with the true velocity. (b) The reconstructed velocity using the separation method based on the estimated single- and multiple-scattering components also demonstrated a linear relation. (c) and (d) show the relative estimation errors of the velocity in the multiple-scattering component for different settings of $Q$ and $S$.}
\end{figure}

\section{Results}

\subsection{Theoretical model validation based on EMC simulation and \textit{in vitro} flow phantom experiment}

The EMC simulations theoretically validated the RMT-based separation and estimation of blood flow. Fig.~\ref{fig:1}(a) shows the direct calculation of the contrast and corresponding estimated velocities from the dynamic speckle images of the single-scattering trajectories, multiple-scattering trajectories and hybrid trajectories during the EMC simulations. Here, the tLASCA algorithm was applied, and the contrast values in the center area covering the vessel were averaged to estimate the mean velocity. All three domains produced good linearity for velocity estimation. Thus, the principle of LSCI is valid for single- and multiple-scattering components in dynamic speckle images. The hybrid contrast demonstrates a higher linearity ($R^2 = 0.953$) than the single ($R^2 = 0.812$) and multiple ($R^2 = 0.886$)components. This is mainly due to the insufficient sampling effects in the single/multiple scattering trajectories used in the calculation (the ratio of the single, multiple and hybrid trajectories is $45\% : 55\% : 100\%$). The proposed RMT-based separation method also provides good linearity for velocity estimation using the single ($R^2 = 0.931$) and multiple ($R^2 = 0.938$) components. The linearity improvements in the RMT-based method are due to the spatial window (window size $S=13$) used in the separation strategy.

\begin{figure}[h!]
\centering\includegraphics[width=0.95\textwidth]{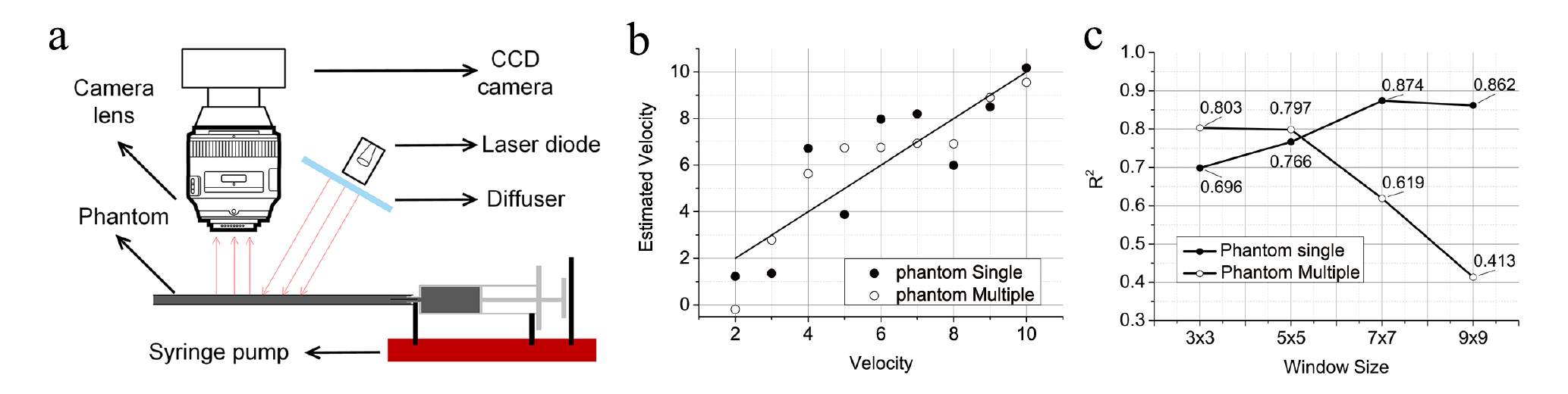}
\caption{\label{fig:2}(a) The imaging setup used in the \textit{in vitro} flow phantom experiment. (b) The estimated velocities from the single- and multiple-scattering components $\emph{v.s.}$ the true velocity in the \textit{in vitro} phantom experiments. (c) Correlation coefficient $R^{2}$ between the estimated velocity and true velocity using different size windows.}
\end{figure}

\textit{in vitro} experiments were performed to further validate the proposed separation strategy. Using a syringe pump, accurate flow control can be achieved (Fig.~\ref{fig:2}(a)). Fig.~\ref{fig:2}(b) shows the estimated velocity and corresponding ground-truth flow velocity based on the single- and multiple-scattering parts in the \textit{in vitro} flow experiments. The proposed method demonstrates high linearity and robustness for velocity estimation for both single scattering $R^{2}=0.766$ and multiple scattering $R^2 = 0.797$ at window size $S=5$. The linearity degradation in the \textit{in vitro} experiment can be attributed to a number of factors: the line width and polarization limitations of the diode laser, noise and distortion in the lens imaging part, velocity spatial distribution in the laminar flow, \emph{etc}. 

\subsection{ Parameter optimization based on EMC simulation and \textit{in vitro} flow phantom experiment}

Mathematically, a larger Wishart RM always provides a better estimation of the spectral density. This can be confirmed by the EMC simulation results. In Fig.~\ref{fig:1}(d), the estimation error for the multiple-scattering component is exponentially decreased to less than $1 \%$ as the window size $S$ linearly increases. It is interesting that linearly increasing the parameter $Q = T / N$ results in damping attenuation of the estimation error for multiple-scattering components (Fig.~\ref{fig:1} (c)). Thus, to achieve sufficient accuracy, $S \geq 3$ and $Q \geq 3$ should be satisfied.

For \textit{in vivo} flow imaging, the spatial diversity of biological tissue leads to a window size S. Thus, N should be constrained in a limited area so that the spectrum of the Wishart RM is not contaminated by inhomogeneous noise. Furthermore, a larger window size decreases the spatial resolution of the reconstructed contrast image.

Fig.~\ref{fig:2}(c) shows the effects of the sampling window size on the linearity of the flow velocity estimation in the \textit{in vitro} flow experiment. An increase in window size ($S=3 \sim 9$) slightly improves the linearity of the flow velocity estimation from the single-scattering component. The linearity of the flow velocity estimation from the multiple-scattering component shows a significant drop at $S=7$ and decreases to $0.290$ at $S=9$. Thus, a large window size is not recommended for realistic applications, especially for the reconstruction of the multiple-scattering components.

Both the EMC simulations and \textit{in vitro} flow experiments suggest an optimized parameter range of $3 \leq S \leq 7$ and $Q \geq 3$. For \textit{in vivo} applications, the parameter $Q$ cannot be arbitrarily large. Because blood flow is constantly changing, it is difficult for longer $T$  to keep the scattering signal ergodic (stationary flow assumption).

\begin{figure}[h!]
\centering\includegraphics[width=0.95\textwidth]{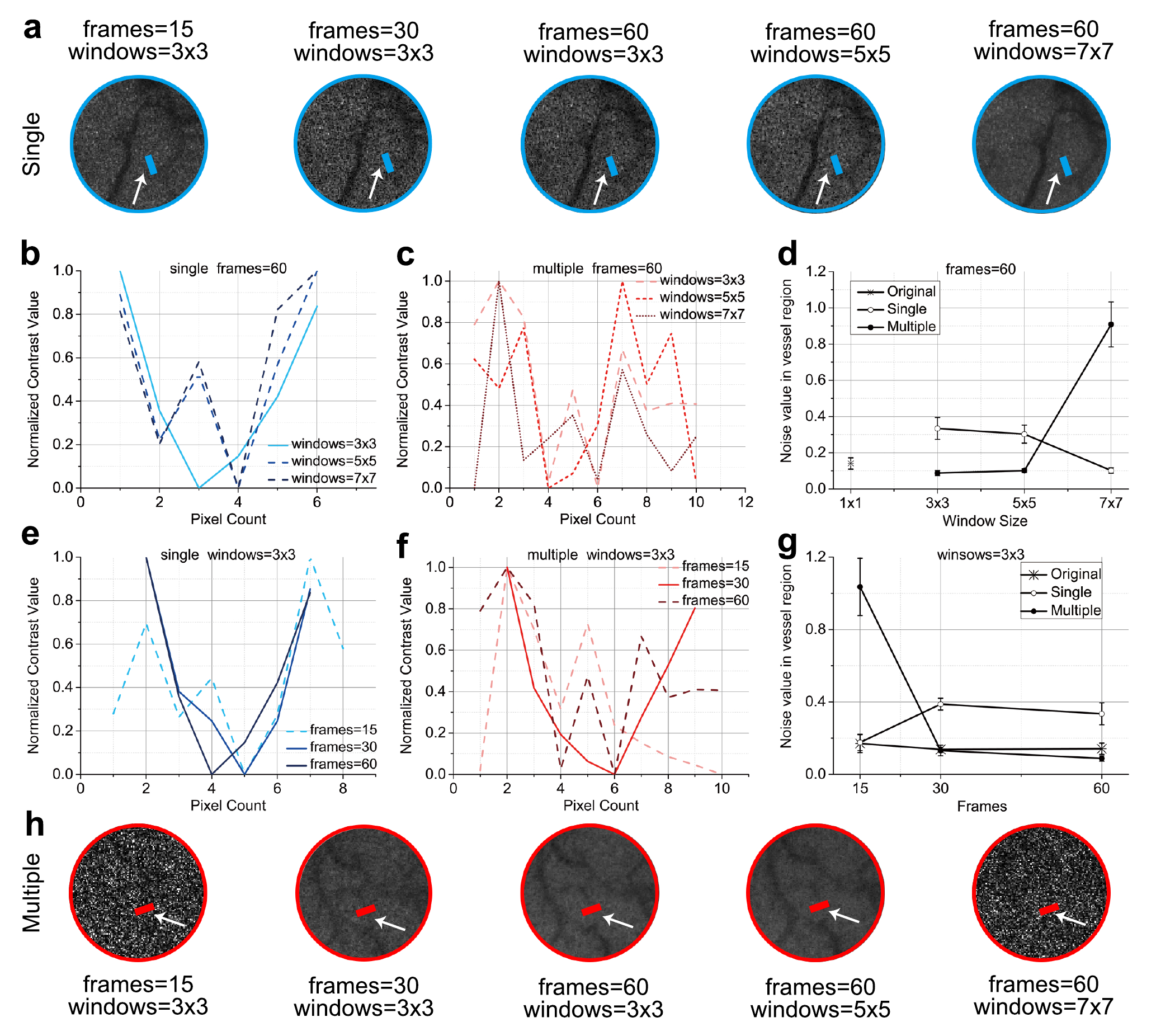}
\caption{\label{fig:3}(a) Enlarged red circle areas reconstructed from the single-scattering part using different numbers of frames and different window sizes. (h) The enlarged red circle areas reconstructed from the multiple-scattering part using different numbers of frames and different window sizes. (b)(e) The contrast values along the cross section of the vessel indicated by the white arrow in (a). (c)(f) The contrast values along the cross section of the vessel indicated by the white arrow in (h). The noise level of the contrast values in the vessel regions calculated by the conventional tLASCA and reconstructed from the single and multiple scattering parts applying different numbers of frames (d) and different window sizes (g).}
\end{figure}

\subsection{Parameter optimization for \textit{in vivo} blood flow imaging}

In this study, using steady-state \textit{in vivo} blood flow imaging experiments, we obtained the final optimized parameter settings. Fig.~\ref{fig:3} (a) and (h) demonstrate the reconstructed single- and multiple-scattering contrast images using different window sizes and $Q$ factors (controlled by the frame amounts). For single scattering, a larger window size and $Q$ value always improve vasculature visibility. Similar effects can be seen for the multiple-scattering contrast images, except for a window size $S = 7$ (the last image in Fig.~\ref{fig:3} (h)). We used noise levels, defined as the standard deviation of the contrast values in the vessel centerline and surrounding areas. The largest vessel was selected to calculate the noise levels in the imaging ROI. 

Fig.~\ref{fig:3} (g) shows the noise levels of various components using different window sizes. The noise level for the hybrid contrast image obtained by directly applying the traditional tLASCA algorithm ($S=1$) is approximately $20 \%$, while the noise level for the multiple-scattering contrast image is $<10 \%$ for $S = 3$ and $S = 5$. For single scattering, a larger spatial window improves the SNR, resulting in lower noise levels. For multiple scattering, the $7 \times 7$ spatial window significantly increases the noise level because of the decay of the estimation linearity. Furthermore, a larger spatial window always degrades spatial resolution and causes velocity distribution distortions. This is demonstrated by the normalized contrast value curves (Fig.~\ref{fig:3} (e) and (f)) along the cross section of the vessels (indicated by the blue arrows in Fig.~\ref{fig:3} (a) and the red arrows in (Fig.~\ref{fig:3} (h))where the standard ‘U’ patterns present when $S=3$ are destroyed when $S = 5$ and $S = 7$. Thus, the $3 \times 3$ spatial window ($N=9$) is the optimal setting for both vasculature visibility and spatial resolution.

Another tuning parameter is $Q = T/N$. Since the optimal $N=9$, the parameter $Q$ is only determined by the number of frames $T$ used in the construction of the Wishart RM. Fig.~\ref{fig:3} (d) shows the noise levels of the hybrid, single-scattering and multiple-scattering contrast images with different $T$ at $S=3$. The single-scattering component has a relatively stable noise level, but it is always larger than that of the hybrid contrast image. The multiple-scattering component has a high noise level when the number of frames is insufficient, $T = 15$. This is because the spectral estimation of the Wishart RM is unstable for $Q = 15 / 9$ (the first image in Fig.~\ref{fig:3} (h)). The noise levels in the multiple-scattering contrast images are even less than those in the hybrid contrast images when a sufficient number of frames are used ($T \geq 30$). An insufficient number of frames also introduces distortions in the normalized contrast value curves (Fig.~\ref{fig:3} (b) and (c)) along the cross section of the selected vessels.

It is critical to optimize the parameter $Q$ with minimal $T$ for functional CBF imaging applications. Although $T=60$ outputs a higher SNR, the temporal resolution is significantly limited, and the ergodic assumption may not be valid. For single scattering, $T=15$ still works well (the first image in Fig.~\ref{fig:3} (a)). For multiple scattering, $T=15$ results in a significant degradation of the contrast image (the first image in Fig.~\ref{fig:3} (h)). Based on the above analysis, T=30 was selected for the following functional CBF imaging experiments.

\subsection{Steady-state CBF imaging and validation by fluorescence imaging}

\begin{figure}[h!]
\centering\includegraphics[width=0.95\textwidth]{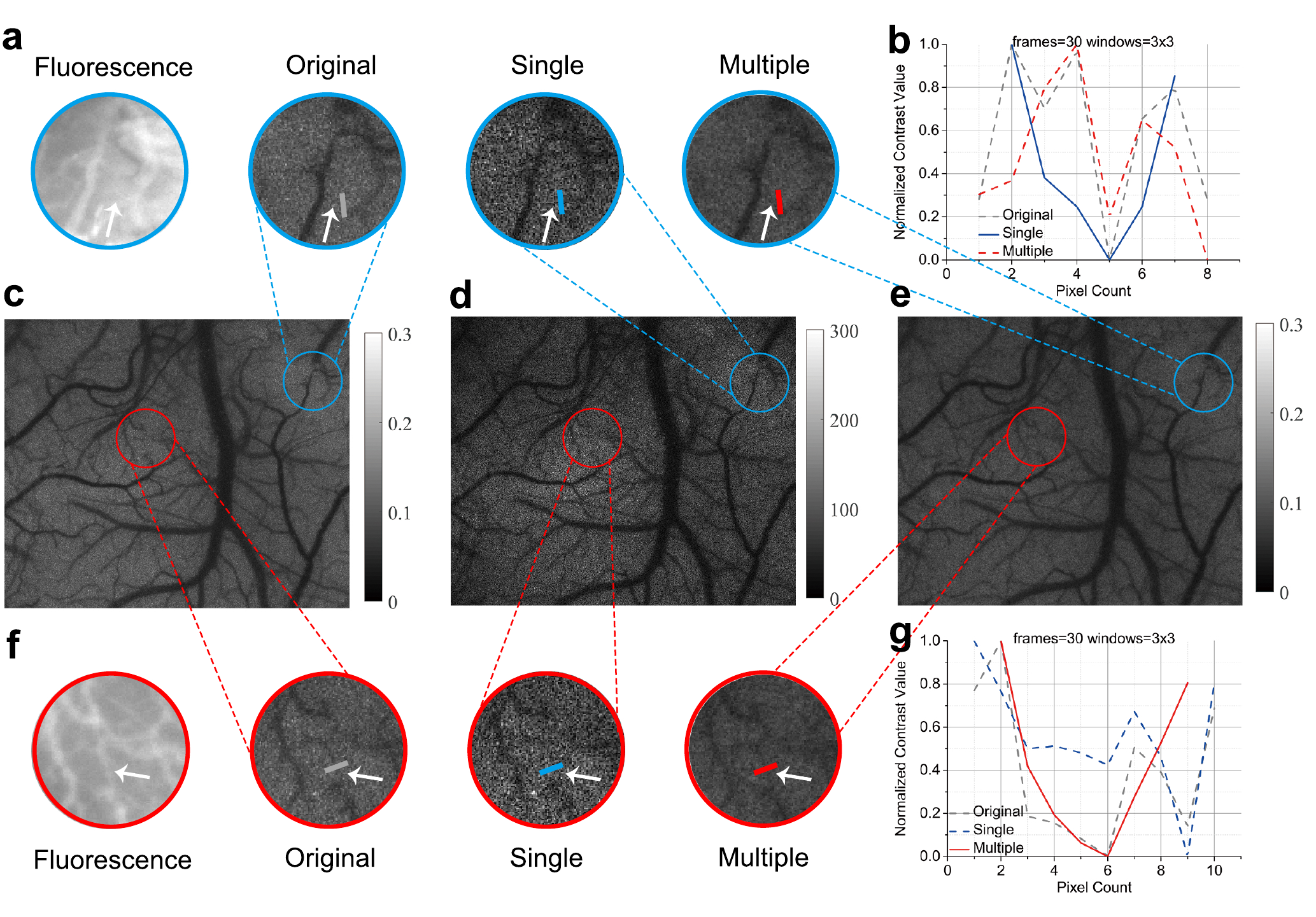}
\caption{\label{fig:4} RMT-based separation was applied in the \textit{in vivo} CBF imaging experiment. (c $\sim$ e) Hybrid, single-scattering and multiple-scattering contrast images; (a) the enlarged blue circle areas in (c $\sim$ e) and corresponding fluorescence image showing the enhancement of the superficial vasculature details in (d). (b) and (g) show the contrast value curves along the cross sections of the selected vessels, as indicated by the white arrows in (a) and (f). (f) shows the enlarged red circle areas in (c $\sim$ e) with more deep tissue vasculature details in (e) than in (c, d). The corresponding fluorescence image confirmed the presence of this vessel.}
\end{figure}

Fig.~\ref{fig:4} (c $\sim$ e) shows the contrast images obtained from the hybrid, single and multiple components, respectively. When the optimal settings ($S=3$ and $T=30$)re used, the multiple-scattering contrast image has the highest imaging SNR. We also identified more deep vasculature details in the multiple-scattering contrast images than in the single and hybrid images, \emph{e.g.}, the vessel indicated by the white arrow in enlarged circle areas (Fig. ~\ref{fig:4} (f)). The existence of this vessel was confirmed by the fluorescence image (the first image in Fig.~\ref{fig:4} (f)) aand the contrast values along the cross-section of the vessel (Fig.~\ref{fig:4} (g)).

The single-scattering contrast image also has more vasculature details in the superficial layers for thicknesses  $<l_t/2$. The white arrow in Fig.~\ref{fig:4} (a) indicates one small vessel that was only visible in the single-scattering contrast image. There is no appearance of such vessels in the hybrid and multiple-scattering contrast images. Even fluorescent imaging cannot reveal this vessel. The existence of this vessel was confirmed by the contrast values along the cross-section of the vessel (Fig.~\ref{fig:4} (b)). Here, we emphasize that both the superficial and deep layers are within the imaging penetration depth.

\subsection{Recovering the functional CBF responses in superficial and deep layers}
\begin{figure}[h!]
\centering\includegraphics[width=1.0\textwidth]{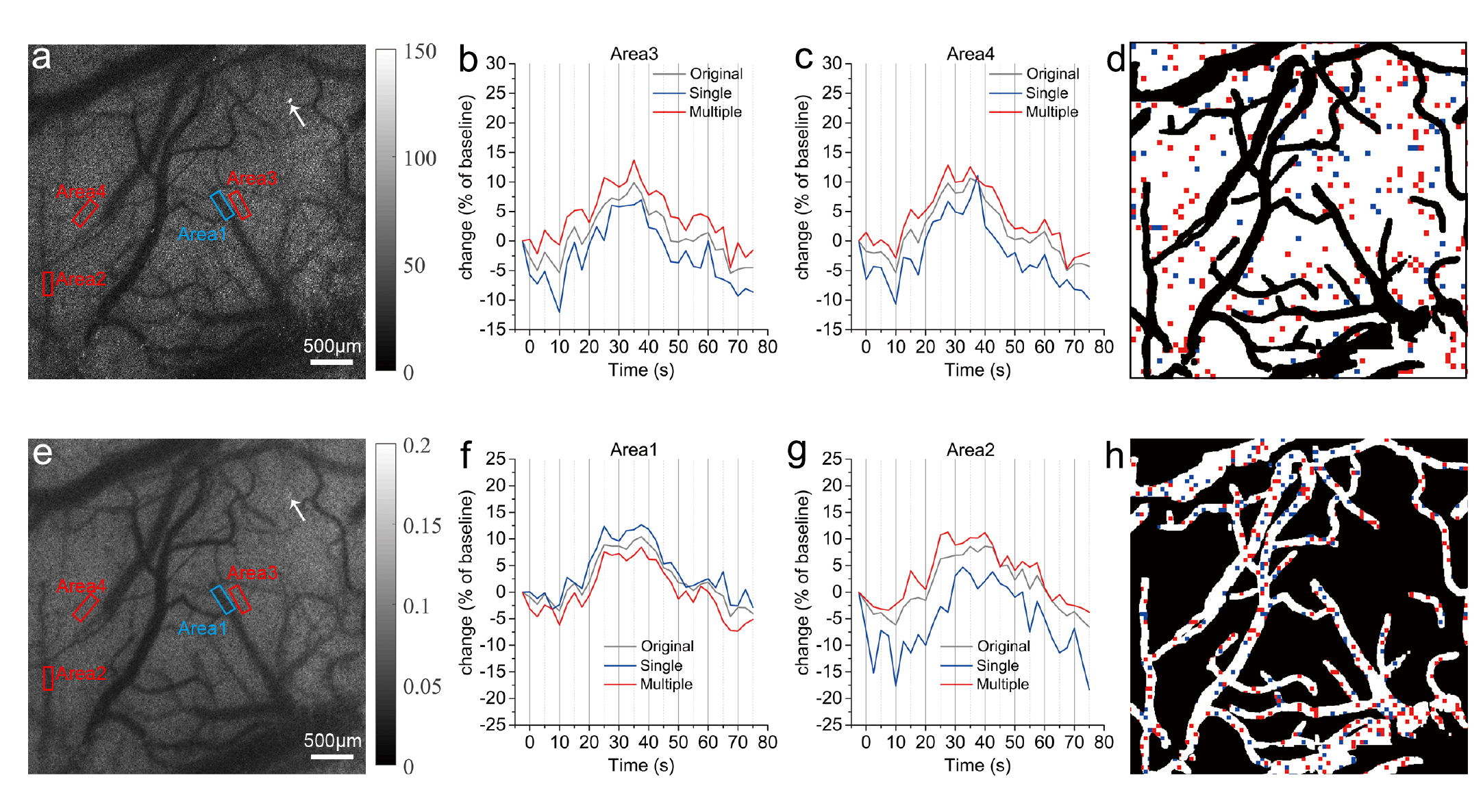}
\caption{\label{fig:5}The proposed strategy is applied to image the functional responses of blood flow and tissue perfusion induced by electrical hind paw stimulation. (a) and (e) Single- and multiple-scattering contrast images. Four typical areas covering the vasculature and tissue are indicated by the red and black boxes in (a) and (e). The corresponding responses of the blood flow and tissue perfusion from the single- and multiple-scattering components are shown in (b), (c), (f) and (g), respectively. The corresponding responses calculated from the original (hybrid) contrast data are also plotted. (d) shows the locations in the tissue area with a significant difference (>10\%) in the average responses of the superficial and deep layers (red dots corresponding to a larger response in the deep layer, blue dots corresponding to a larger response in the superficial layer). (h) shows the locations in the vasculature area with a significant difference (>10\%) in the average responses.}
\end{figure}

Fig.~\ref{fig:5} (a) and (e) shows the single- and multiple-scattering contrast images at their maximum responses after hind paw stimulation using the optimal settings ($S=3$ and $T=30$). In addition to the improved SNR, the multiple-scattering contrast image was also immune to specular noise (white arrow in Fig.~\ref{fig:5} (a)). To analyze the response differences in the superficial and deep layers of the rat cerebral cortex, we show the typical blood flow responses in four rectangular areas ($Area1 \sim Area4$) that cover the vasculature and tissue (indicated by the red and blue boxes in Fig.~\ref{fig:5} (a) and (e)).

Fig.~\ref{fig:5} shows the tissue perfusion responses in the superficial and deep layers induced by hand paw stimulation. The two tissue areas (($Area3$ and $Area4$)) are separated from each other but exhibit similar responses. It is clear that the responses in the deep layers are always larger than those in the superficial layer. To compare the responses of the superficial and deep tissue layers in the full-field ROI, we first segmented the vasculature using an automatic transfer learning-based method and then divided the tissue area into grids with a $7 \times 7$ window size. For each window, the averaged response differences ($0s \sim 70s$) that were larger than the threshold (10\%) were used to identify significant locations. In Fig.~\ref{fig:5} (d), the red points correspond to a larger response in the deep layer, while the blue points correspond to a larger response in the superficial layer. Only 30.64\% of the points were blue, while 69.36\% were red. Therefore, hand paw stimulation-induced brain activities were more significant in deep layers. Previous studies have also confirmed that the coupling between functional stimulation-induced neural activities and tissue perfusion does not usually extend to the superficial layer\cite{1984Mapping, 1998Environmental, 2015Functional}.

Most vasculatures in the imaging ROI have calibers greater than $l_t/2$, indicating that they appear in both the superficial and deep layers within the penetration depth. Different vessels, \emph{e.g.}, $Area1$ and $Area2$, exhibit distinct responses to single- and multiple-scattering components (Fig.~\ref{fig:5} (f) and (g)). For a single vessel, these differences are determined by the depth of the vessel. Vessels with more volume in the superficial layer exhibit larger responses in the single-scattering component. We apply similar comparison processing to demonstrate the significant response differences between the superficial and deep layers of the vessels, as shown in Fig.~\ref{fig:5} (h). 60.49\% of the points were red, while 39.51\% were blue. Some vessel areas had disturbances with both red and blue points. This may reflect non-Newtonian characteristics due to the aggregated motion of red blood cells in \textit{in vivo} CBF\cite{T2011On, Dintenfass1962Thixotropy, 2015Functional}.   

\section{Discussions}

Although the multiple-scattering contrast image reveals more vasculatures, it does not significantly improve the imaging depth. This is mainly due to two reasons. The first is out-of-phase suppression in longer trajectories. The suppression becomes exponentially strengthened in longer paths of multiply-backscattered light. Thus, the fundamental penetration depth limit of LSCI is also applied to the multiple-scattering contrast image. Traditional LSCI outputs a mixture of the single- and multiple-scattering components, with the single-scattering component obscuring the vasculature visualization in the multiple-scattering component. The other reason is absorption effects in biological tissue. Absorption eliminates some light paths, which is more noticeable in longer trajectories. This further exacerbated the limited improvement in the imaging depth. Furthermore, the contrast image from the multiple-scattering component demonstrates blurring effects. More sophisticated deblurring methods are needed to further improve the blood flow imaging of the deep layer.

The separation of blood flow signals in the superficial and deep layers of the sample volume provides a unique opportunity to investigate stimulation-induced blood flow and tissue perfusion. This is especially important for biological tissues with layered structures, \emph{e.g.}, the cerebral cortex and the eye fundus. Responses from different layers may reveal more specific physiological and pathological information. For other imaging modalities that use full-field coherent illumination, our method also provides an easy way to eliminate useless signals from the superficial layer, thus making measurements more accurate. For example, backscattered light from human skin always contains a superficial component, which biases measurement results. Usually, a pair of polarizers are used to extract the cross-polarization part, since most superficial components have parallel polarization characteristics\cite{schmitt1992use, groner1999orthogonal}. However, the illumination power must be significantly increased, which may damage biological tissue. Many trajectories from the deeper layer are also eliminated due to polarization detection, which results in insufficient deep tissue sampling. Furthermore, in many clinical applications, \emph{e.g.}, endoscopy and surgical microscopy, polarization detection cannot be applied due to space limitations and laser power constraints.

In this study, a Wishart random matrix was constructed based on independently-captured dynamic speckle images. The imaging speed was limited due to temporal sampling. To improve the imaging speed, other construction strategies, \emph{e.g.}, multiple cameras and/or multiple illumination directions, can be applied. Multiple-camera devices have been developed and used in a variety of computer vision applications. Speckle images from different cameras are independent samples of the same ensemble. When combined with multiple illumination directions, the data collection procedure can be significantly accelerated. We suggest that the construction strategy be carefully designed based on existing equipment and application scenarios. Similar verification and optimization procedures, such as EMC simulations and \textit{in vitro} simulations, should be conducted to ensure that the RMT-based method is used correctly.

Although the separation of single and multiple scattering photons is advantageous in LSCI applications, the exact depth information remains unresolved. This is the intrinsic difficulty of LSCI as a $2 \mathrm{D}$ full-field imaging modality. However, the theoretical framework can be applied to other $3 \mathrm{D}$ imaging methods, \emph{e.g.}, multifocal reconstruction, DOT/DCT and OCT, where the separation of single and multiple scattering can be resolved with depth coordinates. The theoretical basis should be re-established based on the specific illumination and detection principles.

\section{Conclusion}

In conclusion, we investigated the spectral density of a Wishart random matrix constructed from dynamic speckle images and proposed a separation strategy for the single-scattering and multiple-scattering components. The corresponding first- and second-order statistics can be estimated. We used electric field Monte Carlo (EMC) simulations and \textit{in vitro} phantom experiments to verify the theoretical model and optimize the parameter settings. The proposed strategy was applied to \textit{in vivo} blood flow imaging to obtain the corresponding single- and multiple-scattering contrast images. The single-scattering contrast image has more superficial vasculature details, while the multiple-scattering contrast image reveals more deep tissue vasculatures. We also demonstrated that the proposed method provides a unique tool for the functional imaging of blood flow and tissue perfusion in distinct superficial and deep layers.

\begin{backmatter}
\bmsection{Funding}
This study is supported by Med-X Research Fund of Shanghai Jiao Tong University (YG2021QN16) and the Shanghai Science and Technology Commission of Shanghai Municipality (Grant No. 19DZ2280300).

\bmsection{Acknowledgments}
We thank Mr. Hang Song for his help in animal experiments.

\bmsection{Disclosures}
The authors declare no conflicts of interest.

\bmsection{Data availability} Data underlying the results presented in this paper are not publicly available at this time but may be obtained from the authors upon reasonable request.

\end{backmatter}

\bibliography{Main}

\end{document}